\documentclass[twocolumn]{revtex4-1}
\usepackage{color}
\usepackage{graphicx}
\usepackage{amssymb}

\begin{document}

\newcommand{\ada}{adiabatic approximation}
\newcommand{\adt}{adiabatic theorem}
\setlength{\unitlength}{1cm}
%from mymacros...
\newcommand\bra[1]{\left\langle#1\right|}
\newcommand\ket[1]{\left|#1\right\rangle}
\newcommand\eqref[1]{(\ref{#1})}
\newcommand\beq{\begin{equation}}
\newcommand\eeq{\end{equation}}
\newcommand\bea{\begin{eqnarray}}
\newcommand\eea{\end{eqnarray}}
\newcommand\beano{\begin{eqnarray*}}
\newcommand\eeano{\end{eqnarray*}}
\newcommand\eps{\epsilon}
\newcommand\ltwid{\mathrel{
 \raise.3ex\hbox{$<$\kern-.75em\lower1ex\hbox{$\sim$}}}}
\newcommand\be{\beta}
\newcommand\la{\lambda}
\newcommand\La{\Lambda}
\newcommand\ga{{\gamma}}
\newcommand\om{{\omega}}
\newcommand\si{{\sigma}}
\newcommand\bfsi{\mbox{\boldmath$\sigma$}}
\newcommand\bfn{{\mathbf n}}
\newcommand{\apri}{{\em a priori}}
\newcommand{\apos}{{\em a posteriori}}

\begin{flushright}
\vskip-0.75cm
UdeM-GPP-TH-11-204\\
\end{flushright}
\bigskip

\title{Optimizing adiabaticity in quantum mechanics}

\author{R.~MacKenzie, M.~Pineault and L.~Renaud-Desjardins}
\affiliation{Physique des particules, Universit\'e de Montr\'eal,
C.P. 6128, Succ. Centre-ville, Montr\'eal, QC H3C 3J7}

\date{\today}

\begin{abstract}
A condition on the Hamiltonian of a time-dependent quantum mechanical system is derived which, if satisfied, implies optimal adiabaticity (defined below). The condition is expressed in terms of the Hamiltonian and in terms of the evolution operator related to it. Since the latter depends in a complicated way on the Hamiltonian, it is not yet clear how the condition can be used to extract useful information about the optimal Hamiltonian. The condition is tested on an exactly-soluble time-dependent problem (a spin in a magnetic field), where perfectly adiabatic evolution can be easily identified.

\end{abstract}
\maketitle

\section{Introduction}

The {\adt} in quantum mechanics was developed when quantum mechanics was still in its infancy \cite{Born:1928}. (See \cite{Kato:1950} for an early English-language reference, and \cite{Hagedorn:2006} for a recent overview.) It is used in virtually every area of quantum physics. The essential idea underlying the theorem is very simple. Let the Hamiltonian be $H(t)$ and define its instantaneous eigenstates $\ket{n(t)}$ and eigenenergies $E_n(t)$ as the solutions to the time-\emph{independent} Schroedinger equation $H(t)\ket{\psi}=E\ket{\psi}$ (where $t$ is viewed as a parameter). The states $\ket{n(t)}$ evolve in time, as does the solution $\ket{\psi(t)}$ of the time-\emph{dependent} Schroedinger equation
\[
i\frac{\partial}{\partial t}\ket{\psi(t)}=H(t)\ket{\psi(t)}.
\]

In general there is no obvious relation between these two evolutions. The adiabatic theorem states that the system follows the instantaneous eigenstates for infinitely slow evolution. More precisely, suppose the system starts in $\ket{n(0)}$ at $t=0$, and that $\ket{n(t)}$ is gapped at all times (that is, $|E_n(t)-E_m(t)|\geq E_0>0$ for all $t$ and for all $m\neq n$). Suppose that we can control the speed with which the Hamiltonian varies by making the replacement $H(t)\to H_\la(t)\equiv H(\la t)$ with the time interval changed accordingly, $[0,T]\to[0,T/\la]$, so that the limit $\la\to0$ corresponds to infinitely slow evolution. Then the {\adt} states that the solution to the Schroedinger equation
\[
i\frac{d}{dt}\ket{\psi_\la(t)}=H_\la(t)\ket{\psi_\la(t)},\quad\ket{\psi_\la(0)}=\ket{n(0)}
\]
satisfies
\[
\lim_{\la\to0}\ket{\psi_\la(T/\la)}=(\mbox{phase})\ket{n(T)} + O(\la).
\]
Thus, as $\la\to0$, the probability for the system to make a transition to a different instantaneous eigenstate goes to zero. We then say that the evolution is adiabatic.

Infinitely slow evolution is time consuming, to say the least, so it is of interest to be able to make a statement about finitely slow evolution. Intuitively, if the evolution is sufficiently slow, the {\ada}
\beq
\label{eq:ada}
\ket{\psi(t)}\simeq(\mbox{phase})\ket{n(t)}
\eeq
should be reasonable.

This begs the question: how slow must the evolution be for \eqref{eq:ada} to be a good approximation? In a situation where adiabatic evolution is desirable, it is obviously important to know just how slowly the system must evolve in order for this to be the case, in order to accomplish the task required as quickly as possible. (In the case of adiabatic quantum computing \cite{Farhi:1996na,Farhi:2000}, for instance, how the slowness required scales with the size of the system being studied enables comparison of this approach to conventional circuit-based quantum computing.)

There are two time scales at play: the time scale of the evolution of the Hamiltonian and a scale related to the difference in energies. The former is often estimated to be of order $|\bra{m(t)}\dot n(t)\rangle|^{-1}$; the latter can be written $|E_m-E_n|^{-1}$. Usually, the evolution can be considered slow \cite{Messiah:1962} if
\[
\left| \frac{\bra{m(t)}\dot n(t)\rangle}{E_m-E_n}\right| \ll 1.
\]
But this condition does not guarantee that \eqref{eq:ada} is a good approximation. First, it does not necessarily imply slow evolution \cite{MacKenzie:2006b}. Secondly, even in circumstances where it does imply slow evolution, one cannot conclude that the {\ada} is true indefinitely; rather, it implies that the system will ``escape'' from $\ket{n(t)}$ more slowly than for a ``generic'' (non-slow) evolution. Much work has been done recently studying the {\ada} and attempting to give one or more conditions guaranteeing its validity \cite{Marzlin:2004,Tong:2005,Wu:2005,Duki:2005,Vertesi:2005,Ma:2006,MacKenzie:2006a,Jansen:2006,Tong:2007,MacKenzie:2006b,Rigolin:2008,Zhao:2008,Du:2008,Fujikawa:2008,Comparat:2009,Rezakhani:2009,Amin:2009,Boixo:2010,Tong:2010,Tao:2010,Frasca:2011,Cheung:2011}.

In this paper, our goal will be to devise a criterion which implies that a Hamiltonian is optimal in the sense that the evolution is optimally adiabatic. If $\ket{\psi(0)}=\ket{n(0)}$, then we define the {\em adiabaticity} as
\beq
\label{eq:adiabaticity}
A(t)=\left|\bra{n(t)}\psi(t)\rangle\right|^{2}.
\eeq
$A$ cannot exceed unity, and $A=1$ corresponds to perfectly adiabatic evolution (that is, evolution where the {\ada} is exactly satisfied).

The specific question we will address is as follows. Suppose we are given a physical system described by a time-dependent Hamiltonian, and that the duration of evolution (call it $T$) and the initial and final Hamiltonians ($H(0)$ and $H(T)$) are given. Suppose furthermore that, as above, the system is initially in the state $\ket{n(0)}$. The question is: Can we determine an interpolating Hamiltonian $H(t)$ which maximizes the (final) adiabaticity,
\beq
\label{eq:finaladi}
A[H(t)]=\left|\bra{n(T)}\psi(T)\rangle\right|^{2}\quad?
\eeq
In what follows, we will devise such a criterion, and will test it on an exactly-soluble system. Unfortunately, as will be seen below, the criterion is expressed in terms of the evolution operator. Thus, it is not clear how it can be used on a system for which this operator cannot be determined. In the next section we begin by observing that there is a trivial, but impractical, way of obtaining perfectly adiabatic evolution. We then restrict and formulate the question addressed above. In the following section we explain the optimization procedure which results in out main result, \eqref{eq:mainresult}. Following this, we test the result on what is probably the simplest case of a time-dependent Hamiltonian, one which can be solved exactly: a spin in a rotation magnetic field. We end with a summary and concluding remarks.

\section{Statement of the problem}

Before we begin, note that there is actually a trivial (although impractical) way of attaining perfectly adiabatic evolution \cite{Das:2003}. Suppose $H_s$ (where $s: 0\to1$) is a sufficiently smooth family of Hamiltonians which interpolates between $H(0)$ and $H(T)$ (so that $H_0=H(0)$ and $H_1=H(T)$) and for which the instantaneous eigenstate $\ket{n_s}$ is always gapped. Then the following time-dependent Hamiltonian will give perfectly adiabatic evolution in the limit $\La\to\infty$:
\[
H_\La(t)=\left\{
\begin{array}{rl}
\left(1+\frac{3t}{T}(\La-1)\right)H_0&t:0\to\frac{T}{3}\\
\left(2-\frac{3t}{T}\right)\La H_0 + \left(-1+\frac{3t}{T}\right)\La H_1&t:\frac{T}{3}\to\frac{2T}{3}\\
\left( 3\La-2+\frac{3t}{T}(1-\La)\right)H_1&t:\frac{2T}{3}\to T
\end{array}\right. .
\]
The evolution is divided into three steps. During the first step, the Hamiltonian is simply multiplied by a linear function of time, going from $H_0$ to $\La H_0$; $\ket{n(0)}$ remains an eigenstate of the Hamiltonian so the state of the system only changes by a phase. During the second step, the Hamiltonian evolves (again linearly) from $\La H_0$ to $\La H_1$; as $\La$ goes towards infinity, the probability of making a transition from $\ket{n(t)}$ to a different state drops to zero, so in this limit the state of the system at the end of this stage is a phase times $\ket{n(T)}$. During the third step, the Hamiltonian is again multiplied by a linear function of time, going from $\La H_1$ to $H_1$, and as in the first stage the state of the system only changes by a phase.

Suppose we restrict ourselves to situations where the energy eigenvalues are constant, $E_n(t)=E_n$; this eliminates the trivial way just described, and simplifies the analysis to follow. The Hamiltonian then evolves via a unitary transformation
\[
H(t)=V(t)H(0)V^\dagger(t),
\]
where $V^\dagger(t)V(t)=1$. We can assume $V(0)=1$, and the final value $V(T)$ is fixed since the final Hamiltonian is presumed to have been specified in advance; the instantaneous eigenstates are $\ket{n(t)}=V(t)\ket{n(0)}$. We write $A[H(t)]\to A[V(t)]$.

Let the time evolution operator associated with $H(t)$ be $U(t)$, so the state is given by $\ket{\psi(t)}=U(t)\ket{n(0)}$. Then \eqref{eq:finaladi} becomes
\beq
\label{finaladi2}
%A[V(t)]=\left|\bra{n(0)}V^\dagger(T)U(T)\ket{n(0)}\right|^2,
A[V(t)]=\left|\bra{n(0)}V^\dagger(T)U(T)\ket{n(0)}\right|^2,
\eeq
and we wish to find a condition on $V(t)$ for which the adiabaticity is maximal.
%Eq.~\eqref{finaladi2} has an implicit (and highly non-trivial) dependence on $V(t)$ in $U(T)$.

\section{Optimization}

We will adopt a variational approach to find a condition on $V(t)$. Suppose that the adiabaticity is optimized for a certain matrix $V_0(t)$ (to be determined). Then for any variation of $V(t)$ about $V_0(t)$ which is zero initially and finally (so that the initial and final Hamiltonians are unaffected), $A$ is stationary to first order:
\[
\left.\frac{\delta A}{\delta V(t)}\right|_{V(t)=V_0(t)}=0.
\]
(Of course, solutions to this equation will be {\em local extrema}, not necessarily {\em global minima}, of the adiabaticity.)

We write $H_0(t)=V_0(t)H(0)V_0^\dagger(t)$, with $U_0(t)$ the corresponding evolution operator and $A_0$ the optimal (assumed maximal) adiabaticity.

Now consider a small variation of $V(t)$:
\[
V(t)=(1+i h(t))V_0(t),
\]
where $h(t)=h^\dagger(t)$ (in order for $V(t)$ to be unitary), $h(t)\ll1$ (meaning the matrix elements are much less than 1) and $h(0)=h(T)=0$. Then
\[
\left.\frac{\delta A}{\delta V(t)}\right|_{V(t)=V_0(t)}\to
\left.\frac{\delta A}{\delta h(t)}\right|_{h(t)=0}.
\]
In \eqref{finaladi2}, $V(T)=V_0(T)$ is independent of $h(t)$ (so in what follows we will write $\bra{n(0)}V^\dagger(T)=\bra{n(T)}$), but $U(T)$ depends non-trivially on $h(t)$.

The change $V_0(t)\to V(t)$ will induce a small change in the evolution, so we can write
\[
U(t)=(1+ik(t))U_0(t),
\]
where $k(t)=k^\dagger(t)$, $k(0)=0$ and $k\ll1$. Then
\beano
A &=& \Big| \bra{n(T)}(1+ik(T))U_0(T)\ket{n(0)}\Big|^2\\
&=& A_0 + 2i{\rm Im}\left\{ \bra{n(T)}k(T)U_0(T)\ket{n(0)}\right.\times\\
&&\qquad\qquad\qquad\left.\bra{n(0)}U_0^\dagger(T)\ket{n(T)}   \right\} + O(k^2).
\eeano
That $A$ is stationary implies
\beq
\label{eq:stationary}
{\rm Im}\left\{ \bra{n(T)}k(T)U_0(T)\ket{n(0)}\bra{n(0)}U_0^\dagger(T)\ket{n(T)}   \right\}=0
\eeq
for any $h(t)$.

We must now express $k(T)$ in terms of $h(t)$. We note that for the unperturbed problem
\[
i \frac{d}{dt}U_0(t)=H_0(t)U_0(t),\qquad U_0(0)=1,
\]
while for the perturbed problem
\[
i \frac{d}{dt}U(t)=H(t)U(t),\qquad U(0)=1,
\]
with the Hamiltonians and evolution operators related by 
\beano
H(t)&=&(1+i h(t))H_0(t)(1-i h(t)),\\
U(t)&=&(1+ik(t))U_0(t).
\eeano
Direct substitution and taking linear terms in both $h$ and $k$ yields
\[
i\frac{d}{dt}k(t)=[H_0,k] - [H_0,h],\qquad k(0)=0.
\]
Standard techniques yield the following solution, as can be verified directly:
\[
k(T)=U_0(T)\, i\int_0^T dt\, U_0^\dagger(t) [H_0(t),h(t)] U_0(t) U_0^\dagger(T)
\]
Substituting in \eqref{eq:stationary}, we see that
\[
\begin{array}{l}
{\rm Re}\int_0^T dt\left\{ \bra{n(T)}U_0(T) U_0^\dagger(t) [H_0(t),h(t)] U_0(t) \ket{n(0)}
\right.\\
\left.~~~~~~~~~~~~~~~~~~~~~~~~~~~~\times{ \bra{n(0)}U_0^\dagger(T)\ket{n(T)}  } \right\}=0
\end{array}
\]
for any Hermitian $h(t)$. Each of the factors in the braces can be analyzed semi-intuitively. The second, which can be loosely described as the ``square root" of $A_0$, is simply the amplitude for the optimal evolution to be adiabatic. The first factor, read right to left, is: starting in the initial state, evolving optimally for a time $t$, an effect of the perturbation acting on the state at time $t$, evolving for the remaining time $T-t$, and projecting onto the final instantaneous eigenstate. Thus, roughly speaking, the sum of all possible first-order changes to $A$ must be zero.

We can progress further by analyzing what we mean by the statement ``for any Hermitian $h(t)$." We can write $h(t)=\la_i f_i(t)$ (sum on $i$ implicit), where $\{\la_i\}$ are a basis of Hermitian matrices of the appropriate dimension and $f_i(t)$ are arbitrary functions (except that they are zero initially and finally). We can vary these functions independently, and indeed if one of them is nonzero at one instant and the rest are zero for all times, the above condition becomes
\bea
\label{eq:mainresult}
{\rm Re}\left\{ \bra{n(T)}U_0(T) U_0^\dagger(t) [H_0(t),\la_i] U_0(t) \ket{n(0)}
\right.&&\nonumber\\
\left.~~~~~~~~~~~~~~\times\bra{n(0)}U_0^\dagger(T)\ket{n(T)}   \right\}=0,~~~~~&&
\eea
an equation which must be true for all intermediate times $t$ and for all $i$. This is in principle an equation to be solved for $H_0$. However, as mentioned above, $U_0$ depends in a highly non-trivial way on $H_0$. Thus it is far from obvious how to use it to learn something about a system for which the time evolution operator is unknown (and if it is known, then the adiabatic approximation is of limited use).

\section{Test of the condition}

\begin{figure}[hb]
\begin{centering}
\includegraphics[width=5cm]{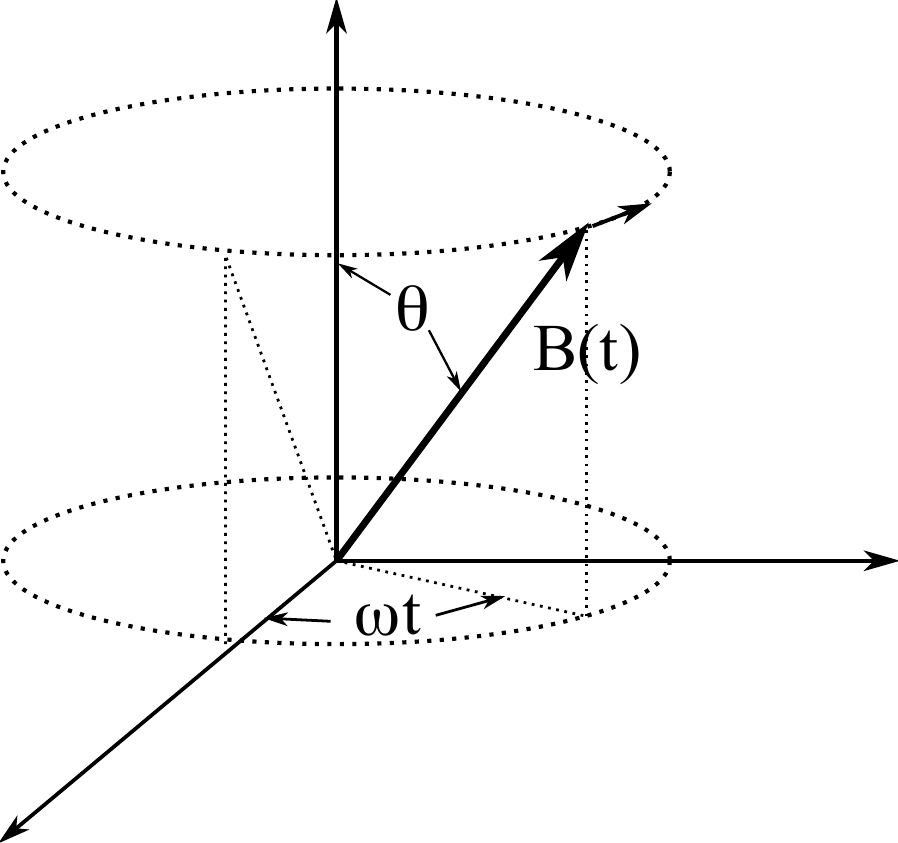}
\par\end{centering}
\caption{Rotating magnetic field with angular frequency $\omega$ and angle with respect to axis of rotation $\theta$.}
\label{fig:one}
\end{figure}

We can at least test the condition on an exactly soluble model: the much-studied case of a spin-1/2 particle in a constantly rotating magnetic field. The geometry is illustrated in Figure \ref{fig:one}.

\begin{figure}[hb]
\begin{centering}
\includegraphics[width=5cm]{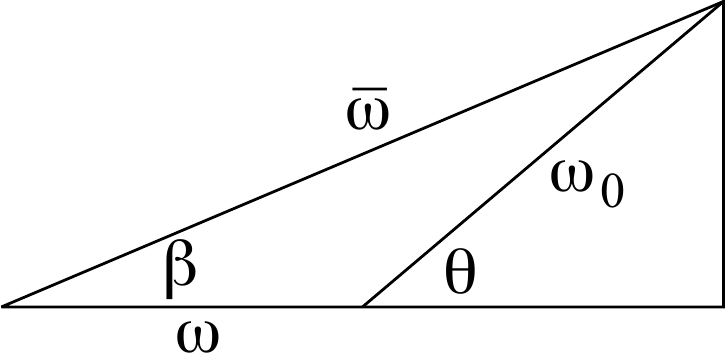}
\par\end{centering}
\caption{Illustration of $\be$, $\bar\om$.}
\label{fig:two}
\end{figure}

The Hamiltonian is
\bea
\label{eq:ham}
H(t)&=&-\frac{\om_0}2\hat\bfn(t)\cdot\bfsi\nonumber\\
&=&-\frac{\om_0}2 e^{-i\frac{\om t}2\si_3} e^{-i\frac{\theta}2\si_2}\si_3
e^{i\frac{\theta}2\si_2} e^{i\frac{\om t}2\si_3}
\eea
where $\hat\bfn$ is a unit vector in the direction of the magnetic field and $\om_0$ is the Larmor frequency.
The time evolution operator is
\beq
\label{eq:timeevolution}
U(t)=e^{-i\frac{\om t}2\si_3} e^{i\frac{\bar\om t}2(c_\be\si_3 + s_\be\si_1)}
\eeq
where $\be$, $\bar\om$ are defined in Figure \ref{fig:two}.

Since the model can be solved exactly, the adiabaticity can be calculated analytically as a function of $t$; the result (supposing we have started in an eigenstate of the initial Hamiltonian, e.g. $\exp(-i\theta\si_2/2)\ket{+}$) is
\[
A(t)=1-\left(\frac{\om s_\theta}{\bar\om}\right)^2 s_{{\bar\om t}/2}^2
\]
$A$ is obviously maximized when $\bar\om t = 2\pi m$, where $m\in{\mathbb{Z}}$.

Thus, we can imagine being asked to find a Hamiltonian which optimizes the adiabaticity for the following parameters of the problem:
\begin{itemize}
\item system: spin 1/2
\item $H(0)=-\frac{\om_0}2 e^{-i\frac{\theta}2\si_2}\si_3 e^{i\frac{\theta}2\si_2}$
\item $H(T)=-\frac{\om_0}2 e^{-i\frac{\om T}2\si_3} e^{-i\frac{\theta}2\si_2}\si_3
e^{i\frac{\theta}2\si_2} e^{i\frac{\om T}2\si_3}$
\item $T=2m\pi/\bar\om$ with $\bar\om$ as defined in Figure \ref{fig:two}.
\end{itemize}
In principle, we would like to use \eqref{eq:mainresult} to determine an optimal Hamiltonian $H_0(t)$. We have not found a way to do so directly, but we can at least check that \eqref{eq:ham} is indeed optimal. This is simply a matter of substituting \eqref{eq:ham} and \eqref{eq:timeevolution} into \eqref{eq:mainresult}. This is somewhat tedious but absolutely straightforward, and we see that indeed for $T=2m\pi/\bar\om$, \eqref{eq:ham} does indeed result in a solution of \eqref{eq:mainresult}, as expected.

\section{Conclusions}

A condition was derived which implies that the adiabaticity (defined in \eqref{eq:adiabaticity}) is maximal.  This condition in principle determines which Hamiltonian or Hamiltonians give rise to perfectly adiabatic evolution. However the condition is expressed in terms of the evolution operator, which depends on a complicated way on the Hamiltonian, so it is not yet clear how to extract useful information about a system for which the evolution operator cannot be determined. We verified that the optimization condition is indeed satisfied when the evolution is optimal, in the case of an exactly-solvable system: a spin in a uniformly rotating magnetic field.

We thank Manu Paranjape for useful discussions. This work was funded in part by the
Natural Science and Engineering Research Council of Canada.

\bibliographystyle{unsrt}
\bibliography{mybibdatabase}

\begin{thebibliography}{10}

\bibitem{Born:1928}
M.~Born and V.~Fock.
\newblock Beweis des adiabatensatzes.
\newblock {\em Z. Phys.}, 51:165--169, 1928.

\bibitem{Kato:1950}
T.~Kato.
\newblock On the adiabatic theorem of quantum mechanics.
\newblock {\em J. Phys. Soc. Japan}, 5:435, 1950.

\bibitem{Hagedorn:2006}
George Hagedorn.
\newblock The simplest situation for adiabatic quantum mechanics.
\newblock Talk presented at {\em Mathematical aspects of quantum adiabatic
  approximation}, Perimeter Institute, 2006.

\bibitem{Farhi:1996na}
Edward Farhi and Sam Gutmann.
\newblock Analog analogue of a digital quantum computation.
\newblock {\em Phys. Rev. A}, 57(4):2403--2406, Apr 1998.

\bibitem{Farhi:2000}
E.~Farhi, J.~Goldstone, S.~Gutmann, and M.~Sipser.
\newblock Quantum computation by adiabatic evolution.
\newblock quant-ph/0001106, 2000.

\bibitem{Messiah:1962}
Albert Messiah.
\newblock {\em Quantum Mechanics, vols. 1 and 2}.
\newblock Wiley, 1962.

\bibitem{MacKenzie:2006b}
R.~MacKenzie, A.~Morin-Duchesne, H.~Paquette, and J.~Pinel.
\newblock Validity of the adiabatic approximation in quantum mechanics.
\newblock {\em Phys. Rev. A}, 76(4):044102, Oct 2007.

\bibitem{Marzlin:2004}
Karl-Peter Marzlin and Barry~C. Sanders.
\newblock Inconsistency in the application of the adiabatic theorem.
\newblock {\em Phys. Rev. Lett.}, 93(16):160408, Oct 2004.

\bibitem{Tong:2005}
D.~M. Tong, K.~Singh, L.~C. Kwek, and C.~H. Oh.
\newblock Quantitative conditions do not guarantee the validity of the
  adiabatic approximation.
\newblock {\em Phys. Rev. Lett.}, 95(11):110407, Sep 2005.

\bibitem{Wu:2005}
Zhaoyan Wu and Hui Yang.
\newblock Validity of the quantum adiabatic theorem.
\newblock {\em Phys. Rev. A}, 72:012114, Jul 2005.

\bibitem{Duki:2005}
Solomon Duki, H.~Mathur, and Onuttom Narayan.
\newblock Is the adiabatic approximation inconsistent?
\newblock quant-ph/0510131v2, 2005.

\bibitem{Vertesi:2005}
T.~V{\`e}rtesi and R.~Englman.
\newblock Perturbative analysis of possible failures in the traditional
  adiabatic conditions.
\newblock {\em Physics Letters A}, 353(1):11 -- 18, 2006.

\bibitem{Ma:2006}
Jie Ma, Yongping Zhang, Enge Wang, and Biao Wu.
\newblock Comment ii on ``inconsistency in the application of the adiabatic
  theorem''.
\newblock {\em Phys. Rev. Lett.}, 97(12):128902, Sep 2006.

\bibitem{MacKenzie:2006a}
R.~MacKenzie, E.~Marcotte, and H.~Paquette.
\newblock Perturbative approach to the adiabatic approximation.
\newblock {\em Phys. Rev. A}, 73(4):042104, Apr 2006.

\bibitem{Jansen:2006}
Sabine Jansen, Mary-Beth Ruskai, and Ruedi Seiler.
\newblock Bounds for the adiabatic approximation with applications to quantum
  computation.
\newblock {\em J. Math. Phys.}, 48:102111, 2007.

\bibitem{Tong:2007}
D.~M. Tong, K.~Singh, L.~C. Kwek, and C.~H. Oh.
\newblock Sufficiency criterion for the validity of the adiabatic
  approximation.
\newblock {\em Phys. Rev. Lett.}, 98(15):150402, Apr 2007.

\bibitem{Rigolin:2008}
Gustavo Rigolin, Gerardo Ortiz, and Victor~Hugo Ponce.
\newblock Beyond the quantum adiabatic approximation: Adiabatic perturbation
  theory.
\newblock {\em Phys. Rev. A}, 78:052508, Nov 2008.

\bibitem{Zhao:2008}
Yan Zhao.
\newblock Reexamination of the quantum adiabatic theorem.
\newblock {\em Phys. Rev. A}, 77:032109, Mar 2008.

\bibitem{Du:2008}
Jiangfeng Du, Lingzhi Hu, Ya~Wang, Jianda Wu, Meisheng Zhao, and Dieter Suter.
\newblock Experimental study of the validity of quantitative conditions in the
  quantum adiabatic theorem.
\newblock {\em Phys. Rev. Lett.}, 101:060403, Aug 2008.

\bibitem{Fujikawa:2008}
Kazuo Fujikawa.
\newblock Adiabatic approximation in the second quantized formulation.
\newblock {\em Phys. Rev. D}, 77:045006, Feb 2008.

\bibitem{Comparat:2009}
Daniel Comparat.
\newblock General conditions for quantum adiabatic evolution.
\newblock {\em Phys. Rev. A}, 80:012106, Jul 2009.

\bibitem{Rezakhani:2009}
A.~T. Rezakhani, W.-J. Kuo, A.~Hamma, D.~A. Lidar, and P.~Zanardi.
\newblock Quantum adiabatic brachistochrone.
\newblock {\em Phys. Rev. Lett.}, 103:080502, Aug 2009.

\bibitem{Amin:2009}
M.~H.~S. Amin.
\newblock Consistency of the adiabatic theorem.
\newblock {\em Phys. Rev. Lett.}, 102:220401, Jun 2009.

\bibitem{Boixo:2010}
S.~Boixo and R.~D. Somma.
\newblock Necessary condition for the quantum adiabatic approximation.
\newblock {\em Phys. Rev. A}, 81:032308, Mar 2010.

\bibitem{Tong:2010}
D.~M. Tong.
\newblock Quantitative condition is necessary in guaranteeing the validity of
  the adiabatic approximation.
\newblock {\em Phys. Rev. Lett.}, 104:120401, Mar 2010.

\bibitem{Tao:2010}
Yong Tao.
\newblock Sufficient condition for the validity of quantum adiabatic theorem.
\newblock arXiv:1010.1326v2 [quant-ph], 2010.

\bibitem{Frasca:2011}
Marco Frasca.
\newblock Consistency of the adiabatic theorem and perturbation theory.
\newblock arXiv:1107.4971v1 [quant-ph], 2011.

\bibitem{Cheung:2011}
Donny Cheung, Peter Hoyer, and Nathan Wiebe.
\newblock Improved error bounds for the adiabatic approximation.
\newblock arXiv:1103.4174v1 [quant-ph], 2011.

\bibitem{Das:2003}
Saurya Das, Randy Kobes, and Gabor Kunstatter.
\newblock Energy and efficiency of adiabatic quantum search algorithms.
\newblock {\em Journal of Physics A: Mathematical and General}, 36(11):2839,
  2003.

\end{thebibliography}

\end{document}